\documentclass[aps,prl,twocolumn,groupedaddress]{revtex4-1}
\usepackage{amsmath}
\usepackage{amssymb}
\usepackage{amsfonts}
\usepackage{color}
\usepackage{graphics}
\usepackage[pdftex]{graphicx}
\usepackage[utf8x]{inputenc}
\usepackage[colorlinks=true]{hyperref}

\begin{document}
\title{Motional Ground State Cooling Outside the Lamb-Dicke Regime}
\author{Y. Yu}
\email{yichaoyu@g.harvard.edu}
\author{N. R. Hutzler}
\altaffiliation{Present address: California Institute of Technology, Division of Physics, Mathematics, and Astronomy. Pasadena, CA, 91125}
\author{J. T. Zhang}
\author{L. R. Liu}
\author{J. D. Hood}
\author{T. Rosenband}
\author{K.-K. Ni}
\email{ni@chemistry.harvard.edu}
\affiliation{Department of Chemistry and Chemical Biology, Harvard University, Cambridge, Massachusetts, 02138, USA}
\affiliation{Department of Physics, Harvard University, Cambridge, Massachusetts, 02138, USA}
\affiliation{Harvard-MIT Center for Ultracold Atoms, Cambridge, Massachusetts, 02138, USA}

\date{\today}

\begin{abstract}
  We report Raman sideband cooling of a single sodium atom to its three-dimensional
  motional ground state in an optical tweezer.
  Despite a large Lamb-Dicke parameter, high initial temperature, and
  large differential light shifts between the excited state and the ground state,
  we achieve a ground state population of $93.5(7)$\% after $53$ ms of cooling.
  Our technique includes addressing high-order sidebands,
  where several motional quanta are removed by a single laser pulse, and
  fast modulation of the optical tweezer intensity.
  We demonstrate that Raman sideband cooling to the 3D motional ground state is possible,
  even without tight confinement and low initial temperature.
\end{abstract}

\maketitle

Trapped neutral atoms, assembled in an array of optical tweezers,
are a promising platform to study quantum information and quantum simulations~
\cite{Schlosser2001,Weiss2004,Isenhower2010,Wilk2010,Kaufman2015,Labuhn2016,Murmann2015}.
The inherent single-particle detection and control, combined with tunable interactions,
allow implementations of neutral-atom-based quantum logic gates~\cite{Isenhower2010,Wilk2010},
novel quantum phases~\cite{Labuhn2016}, and single-photon switches~\cite{Dayan2008,Tiecke2014}.
Advances in real-time re-arrangement of optical tweezers enable rapid preparation of atoms
in large and complex geometries~\cite{Barredo2016,Endres2016}.
Quantum motional control of individual
atoms~\cite{Li2012,Kaufman2012,Thompson2013,Liu2017,Robens2017} enable
studies of the atomic Hong-Ou-Mandel effect~\cite{Kaufman2014},
high-fidelity single qubit gates~\cite{Wang2016},
and efficient coupling of single atoms to photonic crystal cavities~\cite{Thompson2013a}.

Extending optical tweezer arrays to include polar molecules will open a range of
new applications that exploit long-lived molecular internal states
and tunable long range interactions~\cite{DeMille2002,Gorshkov2011,Yan2013}.
Molecules could be assembled from atom pairs in optical tweezers~\cite{Liu2017},
or directly loaded from magneto-optical traps
(MOTs)~\cite{Barry2014,Truppe2017SubDoppler,Anderegg2017}.
For either approach, preparing the constituent atoms or molecules in the lowest motional
quantum state is important to realize long coherence times in quantum applications.

Preparation of single atoms in the motional ground state has been achieved in optical tweezers~\cite{Kaufman2012,Thompson2013,Liu2017,Robens2017}
using Raman sideband cooling (RSC)~\cite{Monroe1995,Kerman2000,Han2000}.
However, RSC in these systems takes place in the Lamb-Dicke (LD) regime, where
the spread of the initial wavefunctions is smaller than the reduced wavelength of light
($\lambdabar$) used to address them.
Systems with a large position spread ($z_{\textrm{rms}}$),
either due to small mass or high initial temperature,
fall outside of the LD regime and undergo strong recoil heating during RSC.
In this letter we demonstrate cooling of single sodium atoms in optical tweezers,
initially outside the LD regime ($z_{\textrm{rms}}/\lambdabar=4.0$), to the motional ground state.
We achieve a 3D ground state population of $P_0=93.5(7)$\% by cooling via
high-order sidebands in a carefully optimized cooling sequence.
Our approach is general and applicable to
other systems. Extending RSC beyond the LD regime opens the possibility
for ground state cooling of systems such as light atoms and
molecules that can be laser-cooled.

Our experiment has an overall repetition rate of $2.5$ Hz and
begins by loading a single sodium atom into an optical tweezer from a MOT~\cite{Hutzler2017-LightShifts}.
The tweezer is created by focussing a $700$ nm wavelength laser beam through an NA=$0.55$ objective to an elliptical waist with radii $\{w_{0,x},w_{0,y}\}\approx \{0.70,0.69\}\,\mu$m.
For $45$ mW of optical power, the trapping frequencies along the three axes are
$\{\omega_x,\omega_y,\omega_z\}/2\pi = \{479(4), 492(5), 86(1)\}\ \text{\text{kHz}}$,
where $z$ is the weakly confined axial direction,
and $x$ and $y$ are the tightly confined radial directions.
After each single atom loading attempt, a first image is taken with a $1.5$ ms exposure time
to determine success, which occurs about 50\% of the time due to the loading mechanism~\cite{Schlosser2001}.
During imaging, the atom is cooled via polarization gradient cooling (PGC),
which reduces the temperature of the single atom to $80$ $\mu$K,
corresponding to mean motional states along the three trap axes of
$\{\bar n_x, \bar n_y, \bar n_z\}=\{3.5(3),\, 3.2(3),\, 20(4)\}$, and therefore initial 3D ground state probability of $P_0=0.4$\%.
The atom is then initialized into
the $|F=2, m_F=2\rangle$ stretched state via optical pumping (OP).
Each experimental cycle ends with a second imaging sequence that measures the hyperfine state of the atom as $|F=2\rangle$ or $|F=1\rangle$~\footnote{The final hyperfine state sensitive detection employs a strong beam that is resonant with the $F=2$ to $F'=3$ cycling transition, to heat the $F=2$ population out of the trap. Therefore, any atoms which survive this heat out procedure are interpreted as having been in the $F=1$ state.}. The data analysis includes only those experiments where the first image reveals successful atom loading.  It should be noted that 4~\% of atoms are lost from one image to the next due to recoil heating.

\begin{figure*}
  \includegraphics[height=4.5cm]{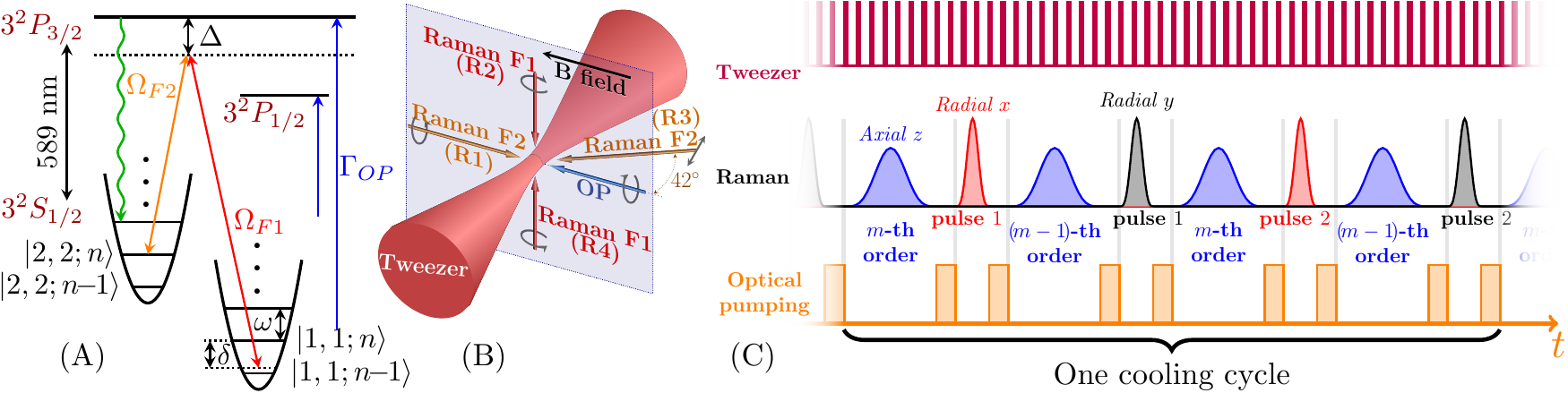}
  \caption{Single Na atom Raman sideband cooling scheme and sequence. (A)
    The Raman transitions between $|2,2;n\rangle$ and $|1,1;n+\Delta n\rangle$ have a one-photon detuning $\Delta=75$ GHz below the $3^2S_{1/2}$ to $3^2P_{3/2}$ transition. Two-photon detuning, $\delta$, is defined relative to the $\Delta n=0$ carrier transition. For optical pumping, we use two $\sigma^+$ polarized transitions, one to pump the atom state out of $|1,1\rangle$ via $3^2P_{3/2}$ and one to pump atoms out of $|2,1\rangle$ via $3^2P_{1/2}$
     to minimize heating of the $|2,2\rangle$ state.
    (B) Geometry and polarizations of the Raman and optical pumping beams relative to the
    optical tweezer and bias magnetic field.  Raman beams R1 and R4 address the radial $x$-mode. R1 and R2 address the radial $y$-mode.  R3 and R4 address the axial $z$-mode, where the beams also couple to radial motion, but this coupling can be neglected when the atoms is cooled to the ground state of motion.
    (C) Schematic of the cooling pulse sequence. The tweezer is strobed at 3 MHz to
    reduce light shifts during optical pumping~\cite{Hutzler2017-LightShifts}.
    Each cooling cycle consists of $8$ sideband pulses.
    The four axial pulses address two sideband orders.
    The two pulses in each radial direction either address $\Delta n=-2$ and $\Delta n=-1$
    or have different durations to drive $\Delta n=-1$, at the end of the cooling sequence when most of the population is below $n=3$.
    The Raman cooling and spectroscopy pulses have Blackman envelopes~\cite{Kasevich1992}
    to reduce off-resonant coupling,
    while the measurement Rabi pulses in Fig.~\ref{f-radial}(B,C) and \ref{f-axial}(B,C)
    have square envelopes to simplify analysis.
    \label{f-setup}}
\end{figure*}

To reduce the temperature of the single atom and
 achieve high ground state population, we apply Raman sideband cooling~\cite{Monroe1995, Kaufman2012}.
The energy levels, cooling sequence, and Raman beam geometries
are shown in Fig.~\ref{f-setup}. RSC consists of two steps:
driving a coherent Raman transition to remove motional quanta, followed by resetting the atom's internal state via OP.
These steps are repeated, until the motional ground state is populated with high probability.

Specifically, Raman transitions are driven in $^{23}$Na between the hyperfine states
$|2, 2\rangle$ and $|1, 1\rangle$ in the presence of an $8.8$ G magnetic field.
The external field is orthogonal to the effective magnetic field of the tweezer,
to reduce vector light shifts~\cite{Kaufman2012,Thompson2013}.
Subsequently, an OP pulse, which consists of two frequencies both with $\sigma^+$-polarization,
is applied to bring the atom out of the $F=1$ manifold and back to $|2, 2\rangle$
via spontaneous emission (Fig.~\ref{f-setup}A).
It is important that the optical pumping step scatter as few photons as possible to minimize recoil heating.
Therefore, to optically pump atoms from $|2, 1\rangle$ into $|2, 2\rangle$
while keeping the target state dark, we apply light resonant with $3^2P_{1/2}$~\cite{Monroe1995, Grobner2017}~\footnote{We find a reduction in the scattering rate by a factor of $130(20)$, as compared to using an OP resonant with $3^2P_{3/2}$, from which the $|2, 2>$ state could always scatter a photon via the excited $|F'=3, m_{F'}=3>$ state.}.

\begin{figure}[b]
  \includegraphics[width=8.5cm]{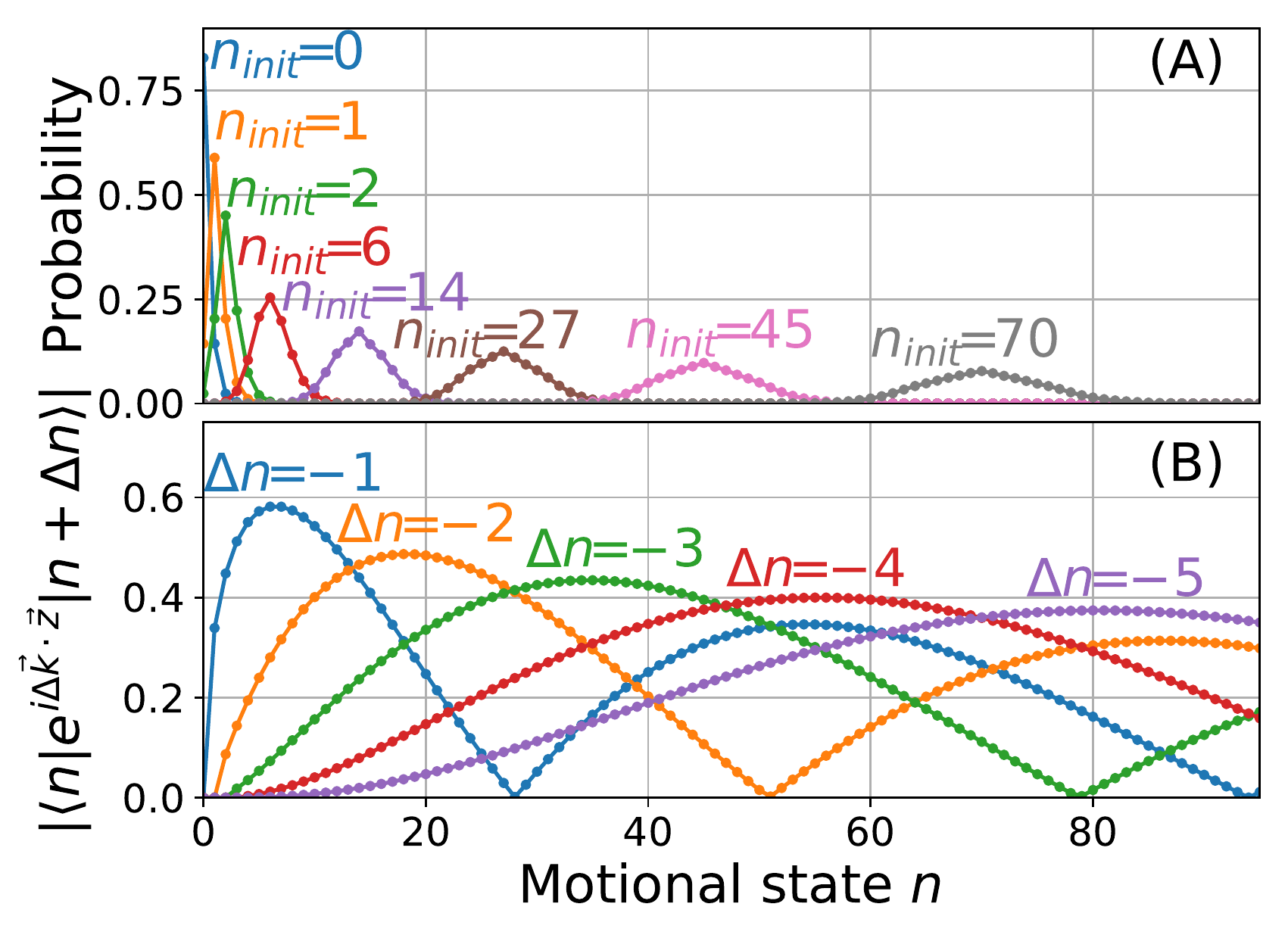}
  \caption{Optical pumping motional-state redistribution and Raman coupling for large LD parameters for the axial direction ($z$). The range plotted covers $95$\% of the initial thermal distribution.
    (A) Motional state distribution after one OP cycle for different initial states motion,
    $n_{\textrm{init}}$. Due to photon-recoil and the large LD parameter, $\eta^{\textrm{OP}}_z=0.55$,
    there is a high probability of $n$ changing.
    (B) Matrix elements for Raman transition deviate from
    $\sqrt{n}$ scaling with multiple minima. During cooling, we utilize the fact that high motional states couple most effectively to sidebands with large $|\Delta n|$.
    \label{f-ld}}
\end{figure}

During optical pumping, photon recoil can lead to significant heating,
especially along the weakly-confined axial direction ($z$).
For an atom starting in an axial motional level $n_{\textrm{init}}$,
the probability to end in another axial state $n$ after absorbing and
re-emitting one photon is approximated by an angular integral of the squared matrix element
$|\langle n_{\textrm{init}}|e^{i \Delta k_{\textrm{OP}} z}| n\rangle|^2$~\cite{ItanoWineland1979}.
Here $\Delta k_{\textrm{OP}}$ is the wave vector difference between absorbed and emitted photons,
projected onto the $z$-axis.
As shown in Fig.~\ref{f-ld}A,
the probability of gaining motional energy during OP grows with $n_{\textrm{init}}$,
and is an important effect for sideband-cooling outside the Lamb-Dicke regime.

Fortunately, the large LD parameters in our system also provide opportunities to overcome OP heating.
The LD parameter for Raman transitions is defined as $\eta^{\textrm{R}}\equiv\Delta k z_0$,
where $\Delta k$ is the wave vector difference
between the two beams that drive the Raman transition.
Here $z_0=\sqrt{\hbar/(2 m \omega)}$ is the ground state wavefunction spread
($m$ is the atomic mass).
The Raman transitions in our configuration have LD parameters of
$\{\eta^{\textrm{R}}_x,\eta^{\textrm{R}}_y,\eta^{\textrm{R}}_z\} = \{0.323(2), 0.319(1), 0.36(1)\}$.
To offset heating from OP initially when $n$ is large, higher-order Raman sidebands
($|\Delta n| > 1$) remove several motional quanta in a single cooling pulse (see Fig.~\ref{f-ld}B).
Because the coupling strengths of different orders do not reach minima
for the same state of motion $n$,
using multiple orders of sidebands avoids accumulations of population
near the coupling minima.

Taking the large motional-state changing heating and cooling sources into account,
it is not immediately clear that ground-state cooling can be achieved.
We therefore use a Monte-Carlo simulation to guide our search~\cite{Dalibard1992}.
In particular, we find that alternating the cooling pulses (Fig.~\ref{f-setup}C) between two
neighboring orders for the axial direction and $\Delta n=-2$ and $\Delta n=-1$
for the radial directions
eliminates the accumulation of population in motional states that have zero Raman coupling,
which would halt the cooling process.
The simulation also indicates that setting the coupling strength of each sideband
to drive a Rabi $\pi$-pulse corresponding to the maximum matrix element motional state
(i.e. the maxima in Fig.~\ref{f-ld}B)  yields efficient cooling, initially.
The efficiency of cooling on higher-order sidebands diminishes
as the atom approaches the ground state, so the final cycles utilize only
the $\Delta n=-1$ sideband while alternating between the three axes.

Guided by the simulation results,
we construct our axial cooling sequence by starting at the two highest
observed sideband orders ($\Delta n=-5$ and $\Delta n=-4$)
and decreasing the orders to the next pair after every $6$ to $15$ cycles.
This process is repeated until $\Delta n=-1$,
and most of the population is in the first few excited states.
We then switch to cooling only on $\Delta n=-1$ with two different pulse lengths
in order to efficiently cool atoms in the few remaining motional states.
Radial cooling is performed similarly to axial cooling with the initial cooling
of $\Delta n=-2$ and $\Delta n=-1$ and switching to $\Delta n=-1$ only after $20$ to $30$ pulses.
This sequence gives good initial cooling performance, which is then used to calibrate experimental
parameters, including the Rabi rates of the Raman beams and optical pumping rates.

There are two additional important challenges in cooling single sodium atoms.
First, the initial temperature populates high motional states,
causing the atoms to sample the anharmonicity of the trap away from the center.
Anharmonicity may be defined as $A_{i,n}=(E_{i,n+1}-E_{i,n})/h - \omega_i/(2\pi)$
for each trap axis $i$, and calculated from the quartic term
of the optical tweezers via perturbation theory.
In the paraxial approximation, we find $A_{i,n}=\frac{-3n\hbar}{4\pi m d_i^2}$,
where $d_i$ equals the beam radius for the radial directions and
$d_z\approx\pi w_{0,x}w_{0,y}/\lambda_{\textrm{trap}}$.
Numerically, $\{A_{x,n},A_{y,n},A_{z,n}\}=\{-1.4, -1.4, -0.16\}n$~kHz.
For thermal states, this broadens and shifts high-order sidebands due to the $n$-dependence of the transitions. To mitigate this, we calibrate the frequency of each sideband order individually, and drive the sidebands with a large Rabi frequency and short pulses to Fourier broaden the spectrum.

\begin{figure*}
  \includegraphics[height=4.2cm]{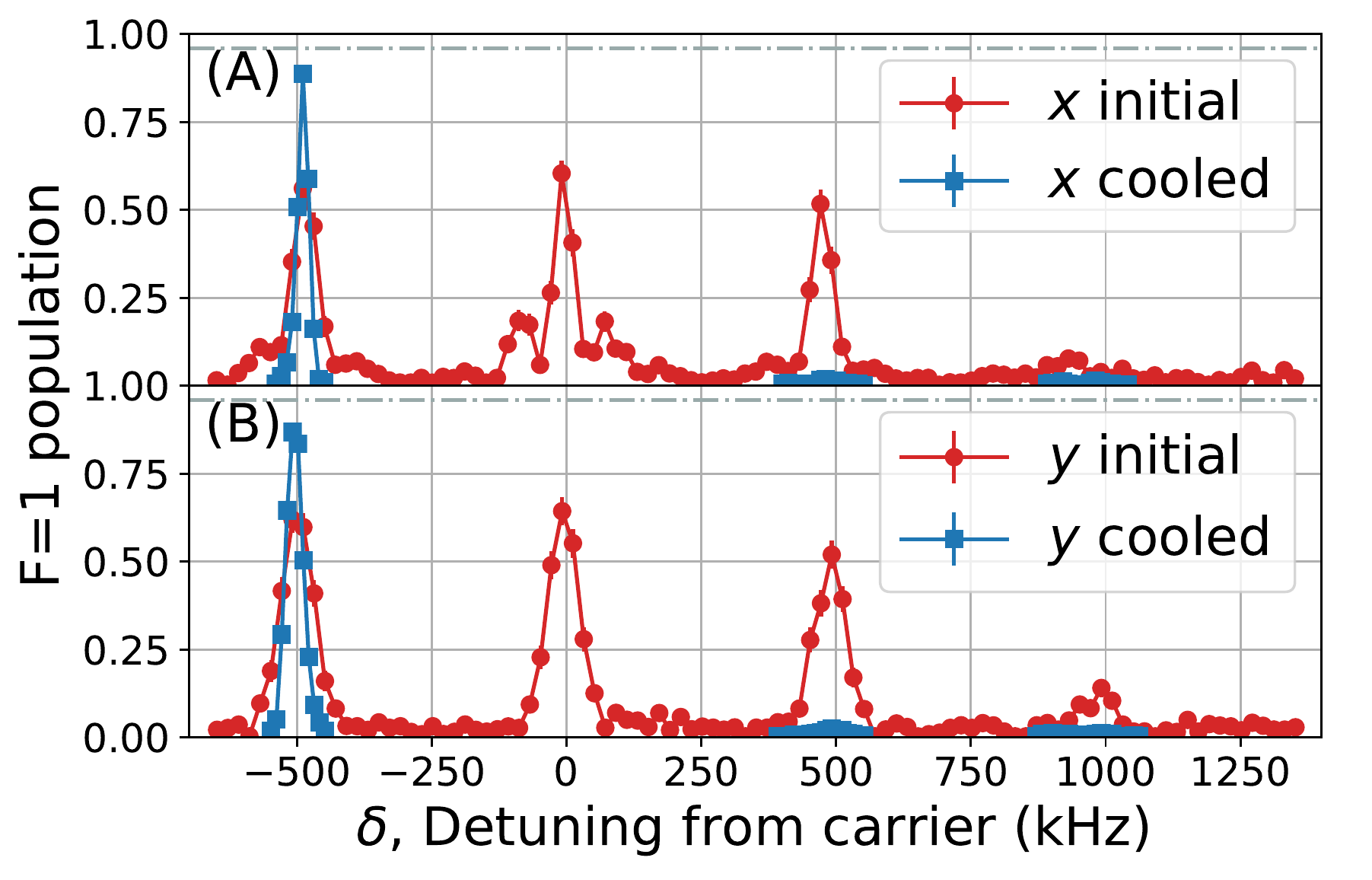}
  \includegraphics[height=4.2cm]{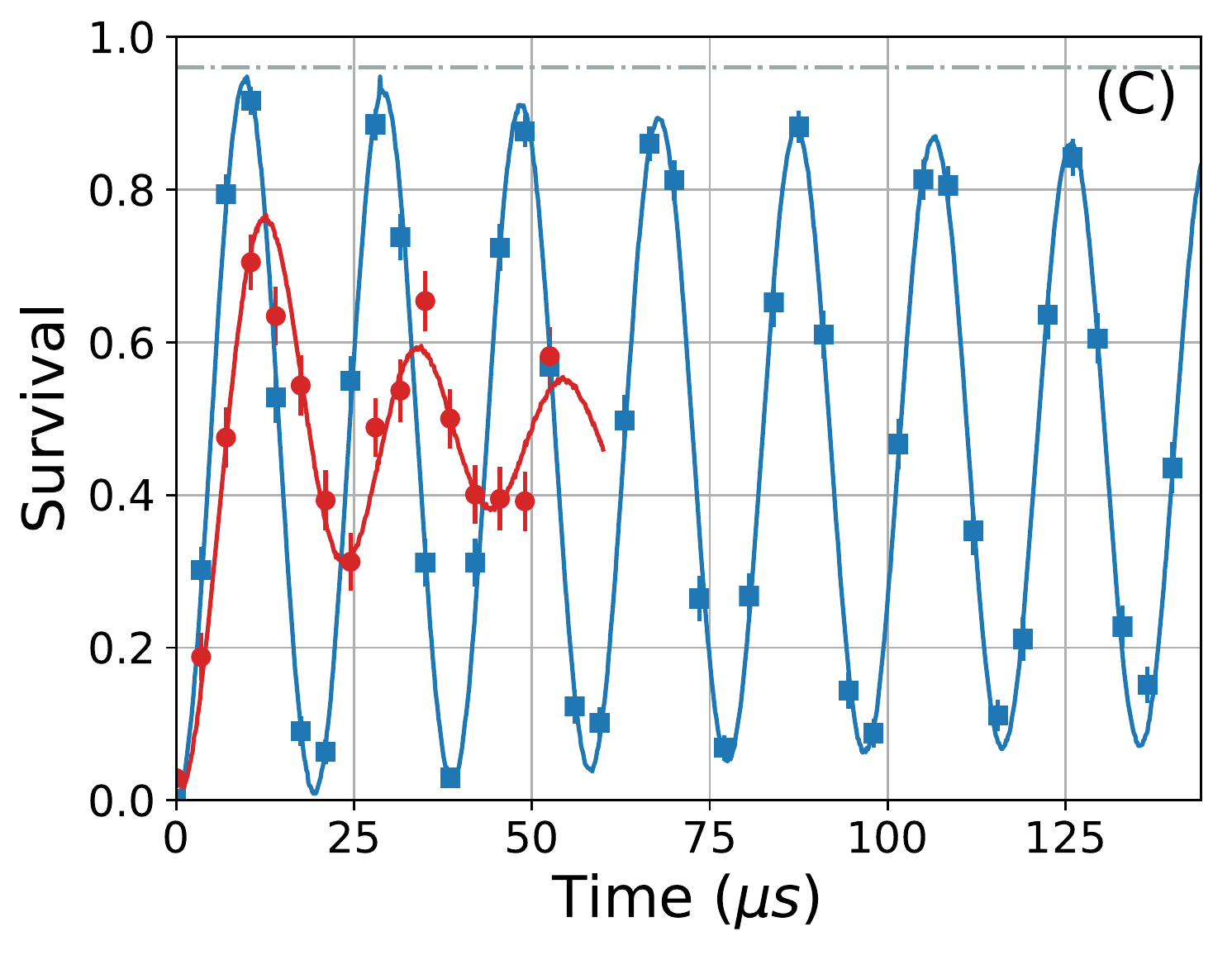}
  \includegraphics[height=4.2cm]{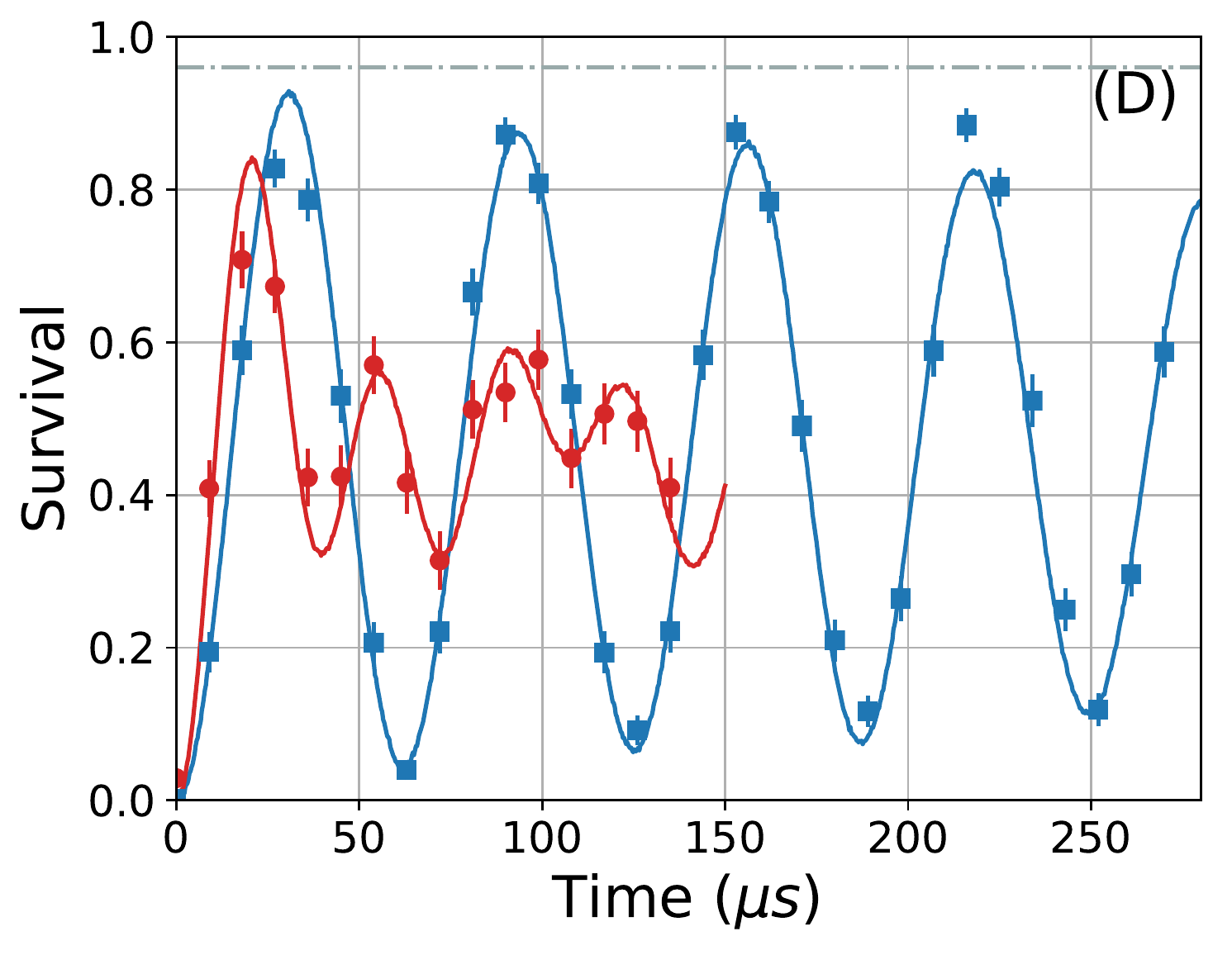}
  \caption{(A,B) Left to right: Radial Raman sideband spectrum of $\Delta n=1,\,\Delta n=0,\,\Delta n=-1,\,\Delta n=-2$ before and after Raman sideband cooling for $x$ and $y$ axis.
    (C,D) Rabi flopping on axis $x$ (B) carrier and (C) $\Delta n_x=1$ sideband
    before (red circle) and after (blue square) RSC.
    Solid lines in (C) and (D) are fits to a Rabi-flopping
    that includes a thermal distribution of motional states~\cite{Meekhof1996}
    as well as off-resonant scattering from the Raman beams.
    The blue lines correspond to a 1D ground state probability of $98.1$\% after cooling
    and the red lines correspond to a thermal distribution of $80$ $\mu$K before RSC.
    The horizontal dashed line in the plot corresponds to the $4\,\%$ probability
    of imaging loss.
    \label{f-radial}}
\end{figure*}
\begin{figure*}
  \includegraphics[height=4.2cm]{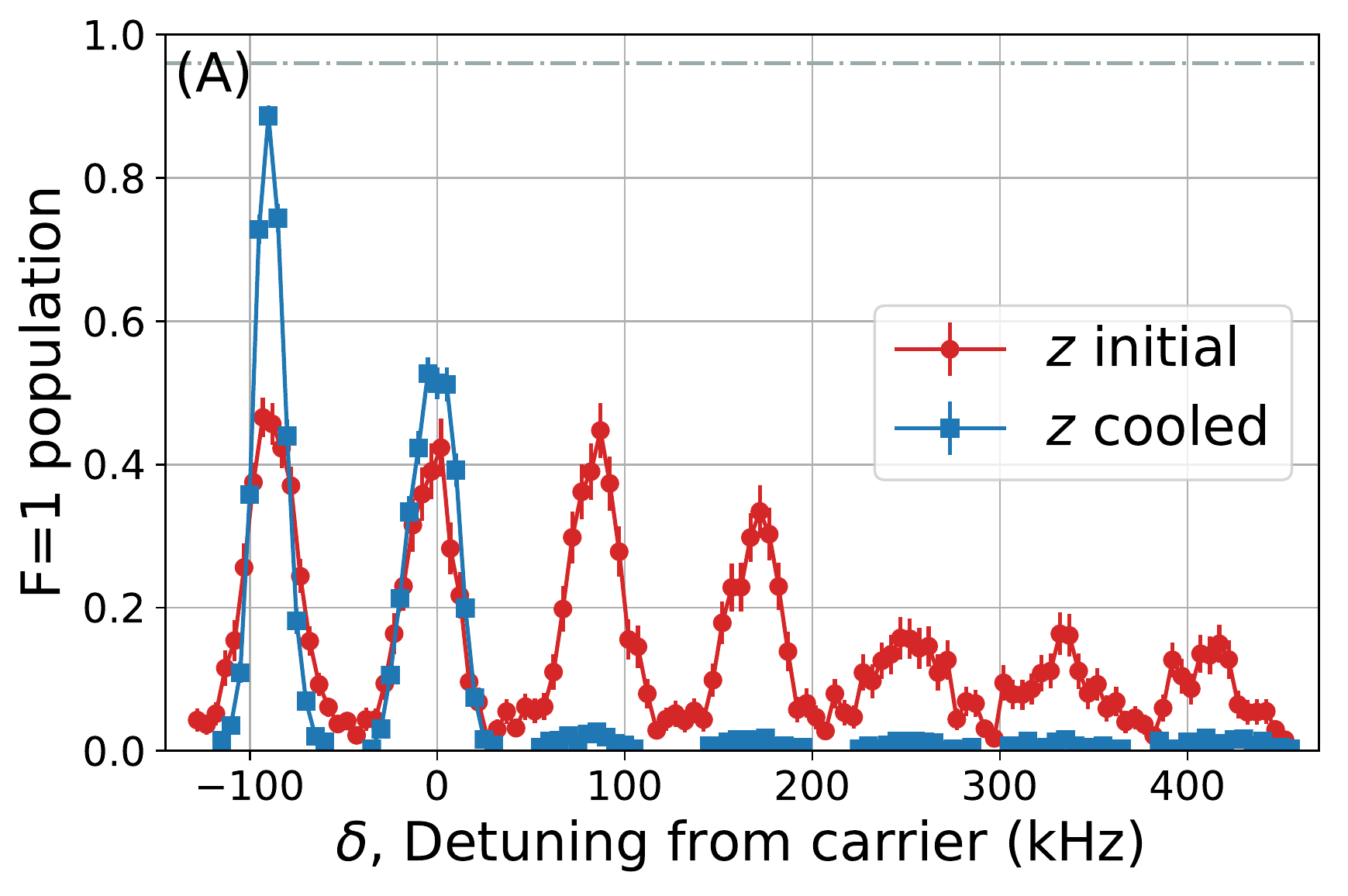}
  \includegraphics[height=4.2cm]{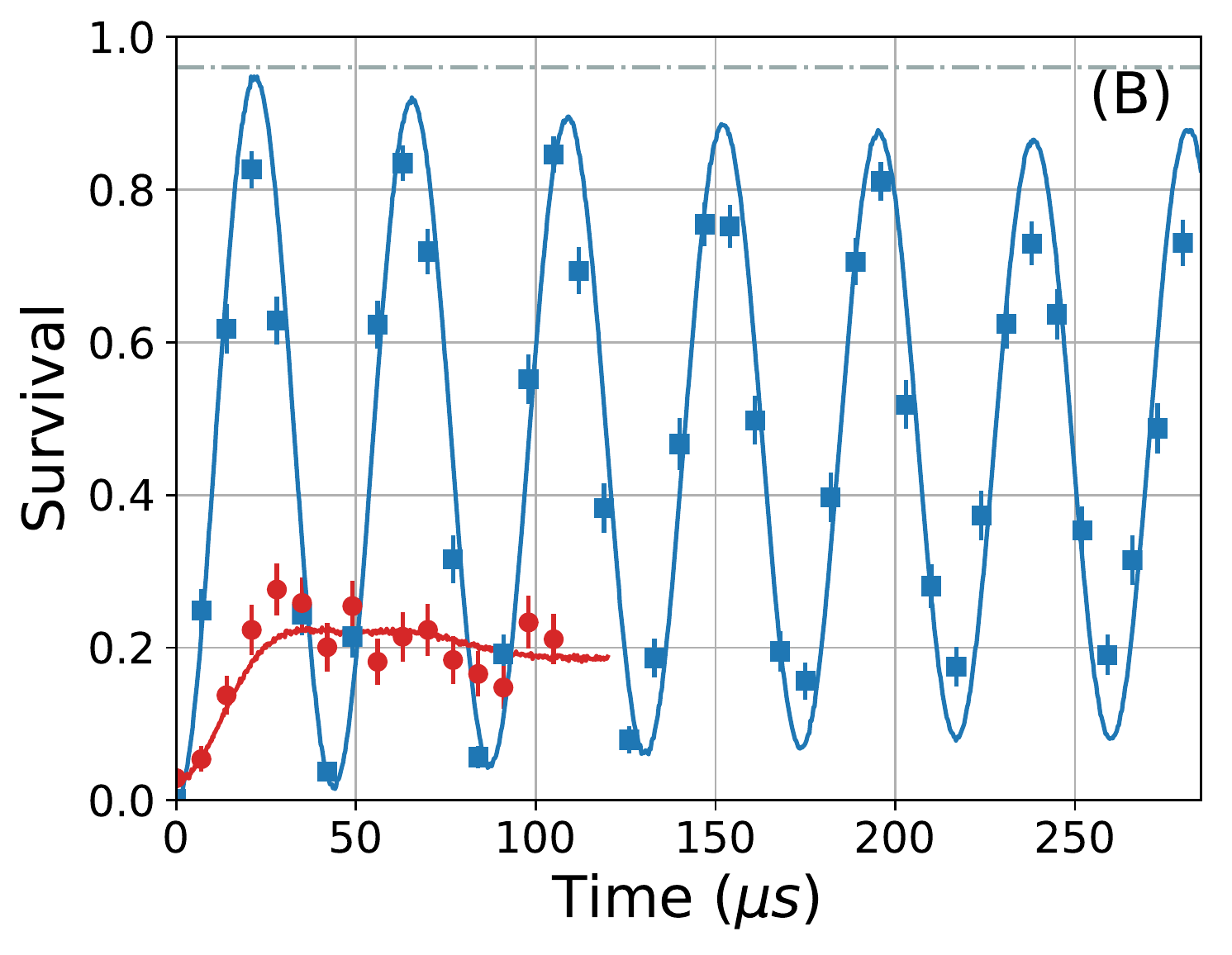}
  \includegraphics[height=4.2cm]{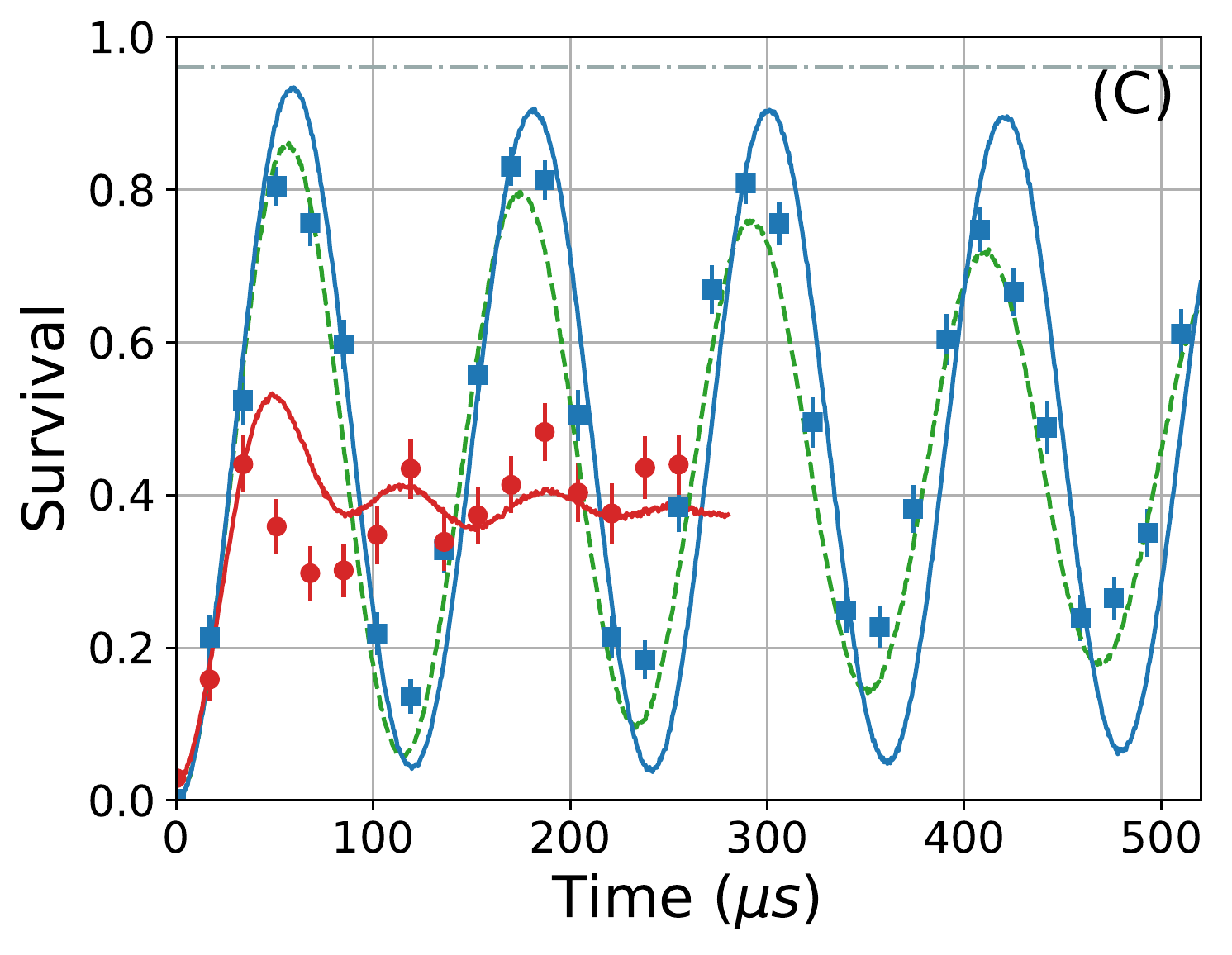}
  \caption{(A) Left to right: Axial Raman sideband spectrum from $\Delta n_z=1,0,-1,\ldots,-5$
    before and after Raman sideband cooling.
    (B,C) Rabi flopping on axial (B) carrier $\Delta n_z=0$  and (C) $\Delta n_z=1$ sideband
    before (red circle) and after (blue square) RSC.
    Solid blue lines in (B) and (C), similar to Fig.~\ref{f-radial}B and C,
    are fits to the Rabi flopping and yield a ground state probability of $95$\%
    after cooling and
    the red lines correspond to a thermal distribution of $80$ $\mu$K before RSC.
    The green dashed line in (C) includes the additional decoherence due to
    a fluctuation of the hyperfine splitting of magnitude $3$ kHz.
    We see that the decoherence effect is strongest for the post-cooling data on
    the $\Delta n_z=1$ sideband where the Rabi frequency is the lowest.
    \label{f-axial}}
\end{figure*}

Second, the tweezers cause a large light shift (as large as $300$ MHz) in the excited state.
We address this by strobing the trapping light at a rate of 3 MHz
during the whole cooling sequence,
similar to our loading and imaging process~\cite{Hutzler2017-LightShifts},
while leaving OP on with constant intensity.
Due to the large light shift, the OP is effectively off whenever the trap light is on.
Since the atom is still addressed by optical pumping light when the trap light is off,
OP remains effective.

Our final cooling results are shown in Fig.~\ref{f-radial} and \ref{f-axial}.
In total, $540$ cooling pulses (total duration $53$ ms) are applied
along three axes with cooling beginning on the radial second order and axial fifth order.
The full RSC pulse sequence is given in~\footnote{See supplemental material}.
To characterize the single atom thermal state before and after cooling,
we perform Raman sideband thermometry~\cite{Monroe1995, Meekhof1996}.
For the more tightly confined radial directions,
we observe clear $\Delta n=1$, $\Delta n=-1$, and $\Delta n=-2$ sidebands before RSC,
as is shown in Fig.~\ref{f-radial}A.
After RSC, the $\Delta n=-1$ and $\Delta n=-2$ sidebands on both radial axes are strongly reduced.
The formula that equates the ratio of $\Delta n=-1$ and $\Delta n=1$ sideband heights to
$\bar{n}/(\bar{n}+1)$ ($\bar{n}$ is the mean motional quantum number of the assumed thermal state)
was derived in the LD regime~\cite{Monroe1995}.
However, it is also valid outside the LD regime.
The resulting $\bar{n}_x=0.019(4)$ and $\bar{n}_y=0.024(3)$ correspond to
ground-state fractions of 98.1(5)\% and 97.6(3)\%,
in agreement with fitted values of 98(1)\% and 95(3)\%
from the Rabi flopping curves~\cite{Meekhof1996} in Fig.~\ref{f-radial}B and \ref{f-radial}C.
The initial temperature of $80\,\mu$K before RSC is obtained from similar fits.

For the weak axial direction, cooling is challenging because the atom starts outside the LD regime.
We observe up to 5th-order Raman cooling sidebands initially,
which indicates population in highly-excited motional states.
Nevertheless, our cooling sequence works efficiently as all the $\Delta n<0$ sidebands are reduced
after RSC (Fig.~\ref{f-axial}A).
Using the ratio of first-order sideband heights, we obtain $\bar{n}_z=0.024(5)$,
which corresponds to a ground state population of
97.6(5)\%, in agreement with a ground state population of 95(4)\% extracted from Rabi flopping
when $\Delta n=0$ (Fig.~\ref{f-axial}B).
For the $\Delta n=1$ sideband (Fig.~\ref{f-axial}C),
we observe additional decoherence that is more pronounced due to the slower Rabi frequency.
The decoherence rate is consistent with magnetic field fluctuations of $1.5$ mG
measured independently in the lab, which would produce a Zeeman shift of $\sim 3$ kHz.

Combining the axial and radial cooling results,
a single Na atom is in the 3D ground state with a probability of 93.5(7)\% after RSC.
The cooling efficiency is limited by spontaneous scattering rate
(0.1-0.2 kHz) from the Raman beams,
as well as spectral broadening from magnetic field fluctuations.

We measure a heating rate that corresponds to decreasing 3D ground state population
at a rate of $\sim0.9$\%/ms.
The rate is consistent with off-resonant scattering of the trapping light~\cite{Grimm2000},
and is predominantly in the axial direction where the trapping beam propagates.

Monte-Carlo simulations show that the ground state probability after RSC
could be enhanced by increasing the detuning of the Raman beams and implementing
better control of the magnetic field. Another improvement could come from
grey molasses cooling, to achieve a lower starting temperature before RSC~\cite{Colzi2016}.

We have shown that despite the difficulty in achieving a low optical cooling temperature
of low mass sodium atoms, three dimensional cooling
with significant ground state population can be achieved by using high-order Raman sidebands
in an optimized cooling sequence.
These techniques are well-suited for a large variety of systems
and open up a route to ground state cooling for other species,
including molecules and exotic atoms.

This work is supported by the Arnold and Mabel Beckman Foundation, the AFOSR Young Investigator Program, the NSF through the CUA,
and the Alfred P. Sloan Foundation.

\bibliography{paper}

\begin{thebibliography}{39}%
\makeatletter
\providecommand \@ifxundefined [1]{%
 \@ifx{#1\undefined}
}%
\providecommand \@ifnum [1]{%
 \ifnum #1\expandafter \@firstoftwo
 \else \expandafter \@secondoftwo
 \fi
}%
\providecommand \@ifx [1]{%
 \ifx #1\expandafter \@firstoftwo
 \else \expandafter \@secondoftwo
 \fi
}%
\providecommand \natexlab [1]{#1}%
\providecommand \enquote  [1]{``#1''}%
\providecommand \bibnamefont  [1]{#1}%
\providecommand \bibfnamefont [1]{#1}%
\providecommand \citenamefont [1]{#1}%
\providecommand \href@noop [0]{\@secondoftwo}%
\providecommand \href [0]{\begingroup \@sanitize@url \@href}%
\providecommand \@href[1]{\@@startlink{#1}\@@href}%
\providecommand \@@href[1]{\endgroup#1\@@endlink}%
\providecommand \@sanitize@url [0]{\catcode `\\12\catcode `\$12\catcode
  `\&12\catcode `\#12\catcode `\^12\catcode `\_12\catcode `\%12\relax}%
\providecommand \@@startlink[1]{}%
\providecommand \@@endlink[0]{}%
\providecommand \url  [0]{\begingroup\@sanitize@url \@url }%
\providecommand \@url [1]{\endgroup\@href {#1}{\urlprefix }}%
\providecommand \urlprefix  [0]{URL }%
\providecommand \Eprint [0]{\href }%
\providecommand \doibase [0]{http://dx.doi.org/}%
\providecommand \selectlanguage [0]{\@gobble}%
\providecommand \bibinfo  [0]{\@secondoftwo}%
\providecommand \bibfield  [0]{\@secondoftwo}%
\providecommand \translation [1]{[#1]}%
\providecommand \BibitemOpen [0]{}%
\providecommand \bibitemStop [0]{}%
\providecommand \bibitemNoStop [0]{.\EOS\space}%
\providecommand \EOS [0]{\spacefactor3000\relax}%
\providecommand \BibitemShut  [1]{\csname bibitem#1\endcsname}%
\let\auto@bib@innerbib\@empty
%</preamble>
\bibitem [{\citenamefont {Schlosser}\ \emph {et~al.}(2001)\citenamefont
  {Schlosser}, \citenamefont {Reymond}, \citenamefont {Protsenko},\ and\
  \citenamefont {Grangier}}]{Schlosser2001}%
  \BibitemOpen
  \bibfield  {author} {\bibinfo {author} {\bibfnamefont {N.}~\bibnamefont
  {Schlosser}}, \bibinfo {author} {\bibfnamefont {G.}~\bibnamefont {Reymond}},
  \bibinfo {author} {\bibfnamefont {I.}~\bibnamefont {Protsenko}}, \ and\
  \bibinfo {author} {\bibfnamefont {P.}~\bibnamefont {Grangier}},\ }\href
  {\doibase 10.1038/35082512} {\bibfield  {journal} {\bibinfo  {journal}
  {Nature}\ }\textbf {\bibinfo {volume} {411}},\ \bibinfo {pages} {1024}
  (\bibinfo {year} {2001})}\BibitemShut {NoStop}%
\bibitem [{\citenamefont {Weiss}\ \emph {et~al.}(2004)\citenamefont {Weiss},
  \citenamefont {Vala}, \citenamefont {Thapliyal}, \citenamefont {Myrgren},
  \citenamefont {Vazirani},\ and\ \citenamefont {Whaley}}]{Weiss2004}%
  \BibitemOpen
  \bibfield  {author} {\bibinfo {author} {\bibfnamefont {D.~S.}\ \bibnamefont
  {Weiss}}, \bibinfo {author} {\bibfnamefont {J.}~\bibnamefont {Vala}},
  \bibinfo {author} {\bibfnamefont {A.~V.}\ \bibnamefont {Thapliyal}}, \bibinfo
  {author} {\bibfnamefont {S.}~\bibnamefont {Myrgren}}, \bibinfo {author}
  {\bibfnamefont {U.}~\bibnamefont {Vazirani}}, \ and\ \bibinfo {author}
  {\bibfnamefont {K.~B.}\ \bibnamefont {Whaley}},\ }\href {\doibase
  10.1103/PhysRevA.70.040302} {\bibfield  {journal} {\bibinfo  {journal} {Phys.
  Rev. A}\ }\textbf {\bibinfo {volume} {70}},\ \bibinfo {pages} {040302}
  (\bibinfo {year} {2004})}\BibitemShut {NoStop}%
\bibitem [{\citenamefont {Isenhower}\ \emph {et~al.}(2010)\citenamefont
  {Isenhower}, \citenamefont {Urban}, \citenamefont {Zhang}, \citenamefont
  {Gill}, \citenamefont {Henage}, \citenamefont {Johnson}, \citenamefont
  {Walker},\ and\ \citenamefont {Saffman}}]{Isenhower2010}%
  \BibitemOpen
  \bibfield  {author} {\bibinfo {author} {\bibfnamefont {L.}~\bibnamefont
  {Isenhower}}, \bibinfo {author} {\bibfnamefont {E.}~\bibnamefont {Urban}},
  \bibinfo {author} {\bibfnamefont {X.~L.}\ \bibnamefont {Zhang}}, \bibinfo
  {author} {\bibfnamefont {A.~T.}\ \bibnamefont {Gill}}, \bibinfo {author}
  {\bibfnamefont {T.}~\bibnamefont {Henage}}, \bibinfo {author} {\bibfnamefont
  {T.~A.}\ \bibnamefont {Johnson}}, \bibinfo {author} {\bibfnamefont {T.~G.}\
  \bibnamefont {Walker}}, \ and\ \bibinfo {author} {\bibfnamefont
  {M.}~\bibnamefont {Saffman}},\ }\href {\doibase
  10.1103/PhysRevLett.104.010503} {\bibfield  {journal} {\bibinfo  {journal}
  {Phys. Rev. Lett.}\ }\textbf {\bibinfo {volume} {104}},\ \bibinfo {pages}
  {010503} (\bibinfo {year} {2010})}\BibitemShut {NoStop}%
\bibitem [{\citenamefont {Wilk}\ \emph {et~al.}(2010)\citenamefont {Wilk},
  \citenamefont {Ga{\"{e}}tan}, \citenamefont {Evellin}, \citenamefont
  {Wolters}, \citenamefont {Miroshnychenko}, \citenamefont {Grangier},\ and\
  \citenamefont {Browaeys}}]{Wilk2010}%
  \BibitemOpen
  \bibfield  {author} {\bibinfo {author} {\bibfnamefont {T.}~\bibnamefont
  {Wilk}}, \bibinfo {author} {\bibfnamefont {A.}~\bibnamefont {Ga{\"{e}}tan}},
  \bibinfo {author} {\bibfnamefont {C.}~\bibnamefont {Evellin}}, \bibinfo
  {author} {\bibfnamefont {J.}~\bibnamefont {Wolters}}, \bibinfo {author}
  {\bibfnamefont {Y.}~\bibnamefont {Miroshnychenko}}, \bibinfo {author}
  {\bibfnamefont {P.}~\bibnamefont {Grangier}}, \ and\ \bibinfo {author}
  {\bibfnamefont {A.}~\bibnamefont {Browaeys}},\ }\href {\doibase
  10.1103/PhysRevLett.104.010502} {\bibfield  {journal} {\bibinfo  {journal}
  {Phys. Rev. Lett.}\ }\textbf {\bibinfo {volume} {104}},\ \bibinfo {pages}
  {010502} (\bibinfo {year} {2010})}\BibitemShut {NoStop}%
\bibitem [{\citenamefont {Kaufman}\ \emph {et~al.}(2015)\citenamefont
  {Kaufman}, \citenamefont {Lester}, \citenamefont {Foss-Feig}, \citenamefont
  {Wall}, \citenamefont {Rey},\ and\ \citenamefont {Regal}}]{Kaufman2015}%
  \BibitemOpen
  \bibfield  {author} {\bibinfo {author} {\bibfnamefont {A.~M.}\ \bibnamefont
  {Kaufman}}, \bibinfo {author} {\bibfnamefont {B.~J.}\ \bibnamefont {Lester}},
  \bibinfo {author} {\bibfnamefont {M.}~\bibnamefont {Foss-Feig}}, \bibinfo
  {author} {\bibfnamefont {M.~L.}\ \bibnamefont {Wall}}, \bibinfo {author}
  {\bibfnamefont {A.~M.}\ \bibnamefont {Rey}}, \ and\ \bibinfo {author}
  {\bibfnamefont {C.~A.}\ \bibnamefont {Regal}},\ }\href {\doibase
  10.1038/nature16073} {\bibfield  {journal} {\bibinfo  {journal} {Nature}\
  }\textbf {\bibinfo {volume} {527}},\ \bibinfo {pages} {208} (\bibinfo {year}
  {2015})}\BibitemShut {NoStop}%
\bibitem [{\citenamefont {Labuhn}\ \emph {et~al.}(2015)\citenamefont {Labuhn},
  \citenamefont {Barredo}, \citenamefont {Ravets}, \citenamefont
  {de~L{\'{e}}s{\'{e}}leuc}, \citenamefont {Macr{\`{i}}}, \citenamefont
  {Lahaye},\ and\ \citenamefont {Browaeys}}]{Labuhn2016}%
  \BibitemOpen
  \bibfield  {author} {\bibinfo {author} {\bibfnamefont {H.}~\bibnamefont
  {Labuhn}}, \bibinfo {author} {\bibfnamefont {D.}~\bibnamefont {Barredo}},
  \bibinfo {author} {\bibfnamefont {S.}~\bibnamefont {Ravets}}, \bibinfo
  {author} {\bibfnamefont {S.}~\bibnamefont {de~L{\'{e}}s{\'{e}}leuc}},
  \bibinfo {author} {\bibfnamefont {T.}~\bibnamefont {Macr{\`{i}}}}, \bibinfo
  {author} {\bibfnamefont {T.}~\bibnamefont {Lahaye}}, \ and\ \bibinfo {author}
  {\bibfnamefont {A.}~\bibnamefont {Browaeys}},\ }\href {\doibase
  10.1038/nature18274} {\bibfield  {journal} {\bibinfo  {journal} {Nature}\
  }\textbf {\bibinfo {volume} {534}},\ \bibinfo {pages} {667} (\bibinfo {year}
  {2015})}\BibitemShut {NoStop}%
\bibitem [{\citenamefont {Murmann}\ \emph {et~al.}(2015)\citenamefont
  {Murmann}, \citenamefont {Bergschneider}, \citenamefont {Klinkhamer},
  \citenamefont {Z\"urn}, \citenamefont {Lompe},\ and\ \citenamefont
  {Jochim}}]{Murmann2015}%
  \BibitemOpen
  \bibfield  {author} {\bibinfo {author} {\bibfnamefont {S.}~\bibnamefont
  {Murmann}}, \bibinfo {author} {\bibfnamefont {A.}~\bibnamefont
  {Bergschneider}}, \bibinfo {author} {\bibfnamefont {V.~M.}\ \bibnamefont
  {Klinkhamer}}, \bibinfo {author} {\bibfnamefont {G.}~\bibnamefont {Z\"urn}},
  \bibinfo {author} {\bibfnamefont {T.}~\bibnamefont {Lompe}}, \ and\ \bibinfo
  {author} {\bibfnamefont {S.}~\bibnamefont {Jochim}},\ }\href {\doibase
  10.1103/PhysRevLett.114.080402} {\bibfield  {journal} {\bibinfo  {journal}
  {Phys. Rev. Lett.}\ }\textbf {\bibinfo {volume} {114}},\ \bibinfo {pages}
  {080402} (\bibinfo {year} {2015})}\BibitemShut {NoStop}%
\bibitem [{\citenamefont {Dayan}\ \emph {et~al.}(2008)\citenamefont {Dayan},
  \citenamefont {Parkins}, \citenamefont {Aoki}, \citenamefont {Ostby},
  \citenamefont {Vahala},\ and\ \citenamefont {Kimble}}]{Dayan2008}%
  \BibitemOpen
  \bibfield  {author} {\bibinfo {author} {\bibfnamefont {B.}~\bibnamefont
  {Dayan}}, \bibinfo {author} {\bibfnamefont {A.~S.}\ \bibnamefont {Parkins}},
  \bibinfo {author} {\bibfnamefont {T.}~\bibnamefont {Aoki}}, \bibinfo {author}
  {\bibfnamefont {E.~P.}\ \bibnamefont {Ostby}}, \bibinfo {author}
  {\bibfnamefont {K.~J.}\ \bibnamefont {Vahala}}, \ and\ \bibinfo {author}
  {\bibfnamefont {H.~J.}\ \bibnamefont {Kimble}},\ }\href {\doibase
  10.1126/science.1152261} {\bibfield  {journal} {\bibinfo  {journal}
  {Science}\ }\textbf {\bibinfo {volume} {319}},\ \bibinfo {pages} {1062}
  (\bibinfo {year} {2008})}\BibitemShut {NoStop}%
\bibitem [{\citenamefont {Tiecke}\ \emph {et~al.}(2014)\citenamefont {Tiecke},
  \citenamefont {Thompson}, \citenamefont {de~Leon}, \citenamefont {Liu},
  \citenamefont {Vuleti{\'{c}}},\ and\ \citenamefont {Lukin}}]{Tiecke2014}%
  \BibitemOpen
  \bibfield  {author} {\bibinfo {author} {\bibfnamefont {T.~G.}\ \bibnamefont
  {Tiecke}}, \bibinfo {author} {\bibfnamefont {J.~D.}\ \bibnamefont
  {Thompson}}, \bibinfo {author} {\bibfnamefont {N.~P.}\ \bibnamefont
  {de~Leon}}, \bibinfo {author} {\bibfnamefont {L.~R.}\ \bibnamefont {Liu}},
  \bibinfo {author} {\bibfnamefont {V.}~\bibnamefont {Vuleti{\'{c}}}}, \ and\
  \bibinfo {author} {\bibfnamefont {M.~D.}\ \bibnamefont {Lukin}},\ }\href
  {\doibase 10.1038/nature13188} {\bibfield  {journal} {\bibinfo  {journal}
  {Nature}\ }\textbf {\bibinfo {volume} {508}},\ \bibinfo {pages} {241}
  (\bibinfo {year} {2014})}\BibitemShut {NoStop}%
\bibitem [{\citenamefont {Barredo}\ \emph {et~al.}(2016)\citenamefont
  {Barredo}, \citenamefont {de~L{\'{e}}s{\'{e}}leuc}, \citenamefont {Lienhard},
  \citenamefont {Lahaye},\ and\ \citenamefont {Browaeys}}]{Barredo2016}%
  \BibitemOpen
  \bibfield  {author} {\bibinfo {author} {\bibfnamefont {D.}~\bibnamefont
  {Barredo}}, \bibinfo {author} {\bibfnamefont {S.}~\bibnamefont
  {de~L{\'{e}}s{\'{e}}leuc}}, \bibinfo {author} {\bibfnamefont
  {V.}~\bibnamefont {Lienhard}}, \bibinfo {author} {\bibfnamefont
  {T.}~\bibnamefont {Lahaye}}, \ and\ \bibinfo {author} {\bibfnamefont
  {A.}~\bibnamefont {Browaeys}},\ }\href {\doibase 10.1126/science.aah3778}
  {\bibfield  {journal} {\bibinfo  {journal} {Science}\ }\textbf {\bibinfo
  {volume} {354}},\ \bibinfo {pages} {1021} (\bibinfo {year}
  {2016})}\BibitemShut {NoStop}%
\bibitem [{\citenamefont {Endres}\ \emph {et~al.}(2016)\citenamefont {Endres},
  \citenamefont {Bernien}, \citenamefont {Keesling}, \citenamefont {Levine},
  \citenamefont {Anschuetz}, \citenamefont {Krajenbrink}, \citenamefont
  {Senko}, \citenamefont {Vuletic}, \citenamefont {Greiner},\ and\
  \citenamefont {Lukin}}]{Endres2016}%
  \BibitemOpen
  \bibfield  {author} {\bibinfo {author} {\bibfnamefont {M.}~\bibnamefont
  {Endres}}, \bibinfo {author} {\bibfnamefont {H.}~\bibnamefont {Bernien}},
  \bibinfo {author} {\bibfnamefont {A.}~\bibnamefont {Keesling}}, \bibinfo
  {author} {\bibfnamefont {H.}~\bibnamefont {Levine}}, \bibinfo {author}
  {\bibfnamefont {E.~R.}\ \bibnamefont {Anschuetz}}, \bibinfo {author}
  {\bibfnamefont {A.}~\bibnamefont {Krajenbrink}}, \bibinfo {author}
  {\bibfnamefont {C.}~\bibnamefont {Senko}}, \bibinfo {author} {\bibfnamefont
  {V.}~\bibnamefont {Vuletic}}, \bibinfo {author} {\bibfnamefont
  {M.}~\bibnamefont {Greiner}}, \ and\ \bibinfo {author} {\bibfnamefont
  {M.~D.}\ \bibnamefont {Lukin}},\ }\href {\doibase 10.1126/science.aah3752}
  {\bibfield  {journal} {\bibinfo  {journal} {Science}\ }\textbf {\bibinfo
  {volume} {354}},\ \bibinfo {pages} {1024} (\bibinfo {year}
  {2016})}\BibitemShut {NoStop}%
\bibitem [{\citenamefont {Li}\ \emph {et~al.}(2012)\citenamefont {Li},
  \citenamefont {Corcovilos}, \citenamefont {Wang},\ and\ \citenamefont
  {Weiss}}]{Li2012}%
  \BibitemOpen
  \bibfield  {author} {\bibinfo {author} {\bibfnamefont {X.}~\bibnamefont
  {Li}}, \bibinfo {author} {\bibfnamefont {T.~A.}\ \bibnamefont {Corcovilos}},
  \bibinfo {author} {\bibfnamefont {Y.}~\bibnamefont {Wang}}, \ and\ \bibinfo
  {author} {\bibfnamefont {D.~S.}\ \bibnamefont {Weiss}},\ }\href {\doibase
  10.1103/PhysRevLett.108.103001} {\bibfield  {journal} {\bibinfo  {journal}
  {Phys. Rev. Lett.}\ }\textbf {\bibinfo {volume} {108}},\ \bibinfo {pages}
  {103001} (\bibinfo {year} {2012})}\BibitemShut {NoStop}%
\bibitem [{\citenamefont {Kaufman}\ \emph {et~al.}(2012)\citenamefont
  {Kaufman}, \citenamefont {Lester},\ and\ \citenamefont
  {Regal}}]{Kaufman2012}%
  \BibitemOpen
  \bibfield  {author} {\bibinfo {author} {\bibfnamefont {A.~M.}\ \bibnamefont
  {Kaufman}}, \bibinfo {author} {\bibfnamefont {B.~J.}\ \bibnamefont {Lester}},
  \ and\ \bibinfo {author} {\bibfnamefont {C.~A.}\ \bibnamefont {Regal}},\
  }\href {\doibase 10.1103/PhysRevX.2.041014} {\bibfield  {journal} {\bibinfo
  {journal} {Phys. Rev. X}\ }\textbf {\bibinfo {volume} {2}},\ \bibinfo {pages}
  {041014} (\bibinfo {year} {2012})}\BibitemShut {NoStop}%
\bibitem [{\citenamefont {Thompson}\ \emph
  {et~al.}(2013{\natexlab{a}})\citenamefont {Thompson}, \citenamefont {Tiecke},
  \citenamefont {Zibrov}, \citenamefont {Vuleti{\'{c}}},\ and\ \citenamefont
  {Lukin}}]{Thompson2013}%
  \BibitemOpen
  \bibfield  {author} {\bibinfo {author} {\bibfnamefont {J.~D.}\ \bibnamefont
  {Thompson}}, \bibinfo {author} {\bibfnamefont {T.~G.}\ \bibnamefont
  {Tiecke}}, \bibinfo {author} {\bibfnamefont {A.~S.}\ \bibnamefont {Zibrov}},
  \bibinfo {author} {\bibfnamefont {V.}~\bibnamefont {Vuleti{\'{c}}}}, \ and\
  \bibinfo {author} {\bibfnamefont {M.~D.}\ \bibnamefont {Lukin}},\ }\href
  {\doibase 10.1103/PhysRevLett.110.133001} {\bibfield  {journal} {\bibinfo
  {journal} {Phys. Rev. Lett.}\ }\textbf {\bibinfo {volume} {110}},\ \bibinfo
  {pages} {133001} (\bibinfo {year} {2013}{\natexlab{a}})}\BibitemShut
  {NoStop}%
\bibitem [{\citenamefont {Liu}\ \emph {et~al.}(2017)\citenamefont {Liu},
  \citenamefont {Zhang}, \citenamefont {Yu}, \citenamefont {Hutzler},
  \citenamefont {Liu}, \citenamefont {Rosenband},\ and\ \citenamefont
  {Ni}}]{Liu2017}%
  \BibitemOpen
  \bibfield  {author} {\bibinfo {author} {\bibfnamefont {L.~R.}\ \bibnamefont
  {Liu}}, \bibinfo {author} {\bibfnamefont {J.~T.}\ \bibnamefont {Zhang}},
  \bibinfo {author} {\bibfnamefont {Y.~Y.}\ \bibnamefont {Yu}}, \bibinfo
  {author} {\bibfnamefont {N.~R.}\ \bibnamefont {Hutzler}}, \bibinfo {author}
  {\bibfnamefont {Y.}~\bibnamefont {Liu}}, \bibinfo {author} {\bibfnamefont
  {T.}~\bibnamefont {Rosenband}}, \ and\ \bibinfo {author} {\bibfnamefont
  {K.-K.}\ \bibnamefont {Ni}},\ }\href {http://arxiv.org/abs/1701.03121}
  {\bibfield  {journal} {\bibinfo  {journal} {arXiv 1701.03121}\ } (\bibinfo
  {year} {2017})}\BibitemShut {NoStop}%
\bibitem [{\citenamefont {Robens}\ \emph {et~al.}(2017)\citenamefont {Robens},
  \citenamefont {Zopes}, \citenamefont {Alt}, \citenamefont {Brakhane},
  \citenamefont {Meschede},\ and\ \citenamefont {Alberti}}]{Robens2017}%
  \BibitemOpen
  \bibfield  {author} {\bibinfo {author} {\bibfnamefont {C.}~\bibnamefont
  {Robens}}, \bibinfo {author} {\bibfnamefont {J.}~\bibnamefont {Zopes}},
  \bibinfo {author} {\bibfnamefont {W.}~\bibnamefont {Alt}}, \bibinfo {author}
  {\bibfnamefont {S.}~\bibnamefont {Brakhane}}, \bibinfo {author}
  {\bibfnamefont {D.}~\bibnamefont {Meschede}}, \ and\ \bibinfo {author}
  {\bibfnamefont {A.}~\bibnamefont {Alberti}},\ }\href {\doibase
  10.1103/PhysRevLett.118.065302} {\bibfield  {journal} {\bibinfo  {journal}
  {Phys. Rev. Lett.}\ }\textbf {\bibinfo {volume} {118}},\ \bibinfo {pages}
  {065302} (\bibinfo {year} {2017})}\BibitemShut {NoStop}%
\bibitem [{\citenamefont {Kaufman}\ \emph {et~al.}(2014)\citenamefont
  {Kaufman}, \citenamefont {Lester}, \citenamefont {Reynolds}, \citenamefont
  {Wall}, \citenamefont {Foss-Feig}, \citenamefont {Hazzard}, \citenamefont
  {Rey},\ and\ \citenamefont {Regal}}]{Kaufman2014}%
  \BibitemOpen
  \bibfield  {author} {\bibinfo {author} {\bibfnamefont {A.~M.}\ \bibnamefont
  {Kaufman}}, \bibinfo {author} {\bibfnamefont {B.~J.}\ \bibnamefont {Lester}},
  \bibinfo {author} {\bibfnamefont {C.~M.}\ \bibnamefont {Reynolds}}, \bibinfo
  {author} {\bibfnamefont {M.~L.}\ \bibnamefont {Wall}}, \bibinfo {author}
  {\bibfnamefont {M.}~\bibnamefont {Foss-Feig}}, \bibinfo {author}
  {\bibfnamefont {K.~R.~A.}\ \bibnamefont {Hazzard}}, \bibinfo {author}
  {\bibfnamefont {A.~M.}\ \bibnamefont {Rey}}, \ and\ \bibinfo {author}
  {\bibfnamefont {C.~A.}\ \bibnamefont {Regal}},\ }\href {\doibase
  10.1126/science.1250057} {\bibfield  {journal} {\bibinfo  {journal}
  {Science}\ }\textbf {\bibinfo {volume} {345}},\ \bibinfo {pages} {306}
  (\bibinfo {year} {2014})}\BibitemShut {NoStop}%
\bibitem [{\citenamefont {Wang}\ \emph {et~al.}(2016)\citenamefont {Wang},
  \citenamefont {Kumar}, \citenamefont {Wu},\ and\ \citenamefont
  {Weiss}}]{Wang2016}%
  \BibitemOpen
  \bibfield  {author} {\bibinfo {author} {\bibfnamefont {Y.}~\bibnamefont
  {Wang}}, \bibinfo {author} {\bibfnamefont {A.}~\bibnamefont {Kumar}},
  \bibinfo {author} {\bibfnamefont {T.-Y.}\ \bibnamefont {Wu}}, \ and\ \bibinfo
  {author} {\bibfnamefont {D.~S.}\ \bibnamefont {Weiss}},\ }\href {\doibase
  10.1126/science.aaf2581} {\bibfield  {journal} {\bibinfo  {journal}
  {Science}\ }\textbf {\bibinfo {volume} {352}},\ \bibinfo {pages} {1562}
  (\bibinfo {year} {2016})}\BibitemShut {NoStop}%
\bibitem [{\citenamefont {Thompson}\ \emph
  {et~al.}(2013{\natexlab{b}})\citenamefont {Thompson}, \citenamefont {Tiecke},
  \citenamefont {de~Leon}, \citenamefont {Feist}, \citenamefont {Akimov},
  \citenamefont {Gullans}, \citenamefont {Zibrov}, \citenamefont {Vuletic},\
  and\ \citenamefont {Lukin}}]{Thompson2013a}%
  \BibitemOpen
  \bibfield  {author} {\bibinfo {author} {\bibfnamefont {J.~D.}\ \bibnamefont
  {Thompson}}, \bibinfo {author} {\bibfnamefont {T.~G.}\ \bibnamefont
  {Tiecke}}, \bibinfo {author} {\bibfnamefont {N.~P.}\ \bibnamefont {de~Leon}},
  \bibinfo {author} {\bibfnamefont {J.}~\bibnamefont {Feist}}, \bibinfo
  {author} {\bibfnamefont {A.~V.}\ \bibnamefont {Akimov}}, \bibinfo {author}
  {\bibfnamefont {M.}~\bibnamefont {Gullans}}, \bibinfo {author} {\bibfnamefont
  {A.~S.}\ \bibnamefont {Zibrov}}, \bibinfo {author} {\bibfnamefont
  {V.}~\bibnamefont {Vuletic}}, \ and\ \bibinfo {author} {\bibfnamefont
  {M.~D.}\ \bibnamefont {Lukin}},\ }\href {\doibase 10.1126/science.1237125}
  {\bibfield  {journal} {\bibinfo  {journal} {Science}\ }\textbf {\bibinfo
  {volume} {340}},\ \bibinfo {pages} {1202} (\bibinfo {year}
  {2013}{\natexlab{b}})}\BibitemShut {NoStop}%
\bibitem [{\citenamefont {DeMille}(2002)}]{DeMille2002}%
  \BibitemOpen
  \bibfield  {author} {\bibinfo {author} {\bibfnamefont {D.}~\bibnamefont
  {DeMille}},\ }\href {\doibase 10.1103/PhysRevLett.88.067901} {\bibfield
  {journal} {\bibinfo  {journal} {Phys. Rev. Lett.}\ }\textbf {\bibinfo
  {volume} {88}},\ \bibinfo {pages} {067901} (\bibinfo {year}
  {2002})}\BibitemShut {NoStop}%
\bibitem [{\citenamefont {Gorshkov}\ \emph {et~al.}(2011)\citenamefont
  {Gorshkov}, \citenamefont {Manmana}, \citenamefont {Chen}, \citenamefont
  {Demler}, \citenamefont {Lukin},\ and\ \citenamefont {Rey}}]{Gorshkov2011}%
  \BibitemOpen
  \bibfield  {author} {\bibinfo {author} {\bibfnamefont {A.~V.}\ \bibnamefont
  {Gorshkov}}, \bibinfo {author} {\bibfnamefont {S.~R.}\ \bibnamefont
  {Manmana}}, \bibinfo {author} {\bibfnamefont {G.}~\bibnamefont {Chen}},
  \bibinfo {author} {\bibfnamefont {E.}~\bibnamefont {Demler}}, \bibinfo
  {author} {\bibfnamefont {M.~D.}\ \bibnamefont {Lukin}}, \ and\ \bibinfo
  {author} {\bibfnamefont {A.~M.}\ \bibnamefont {Rey}},\ }\href {\doibase
  10.1103/PhysRevA.84.033619} {\bibfield  {journal} {\bibinfo  {journal} {Phys.
  Rev. A}\ }\textbf {\bibinfo {volume} {84}},\ \bibinfo {pages} {033619}
  (\bibinfo {year} {2011})}\BibitemShut {NoStop}%
\bibitem [{\citenamefont {Yan}\ \emph {et~al.}(2013)\citenamefont {Yan},
  \citenamefont {Moses}, \citenamefont {Gadway}, \citenamefont {Covey},
  \citenamefont {Hazzard}, \citenamefont {Rey}, \citenamefont {Jin},\ and\
  \citenamefont {Ye}}]{Yan2013}%
  \BibitemOpen
  \bibfield  {author} {\bibinfo {author} {\bibfnamefont {B.}~\bibnamefont
  {Yan}}, \bibinfo {author} {\bibfnamefont {S.~A.}\ \bibnamefont {Moses}},
  \bibinfo {author} {\bibfnamefont {B.}~\bibnamefont {Gadway}}, \bibinfo
  {author} {\bibfnamefont {J.~P.}\ \bibnamefont {Covey}}, \bibinfo {author}
  {\bibfnamefont {K.~R.~A.}\ \bibnamefont {Hazzard}}, \bibinfo {author}
  {\bibfnamefont {A.~M.}\ \bibnamefont {Rey}}, \bibinfo {author} {\bibfnamefont
  {D.~S.}\ \bibnamefont {Jin}}, \ and\ \bibinfo {author} {\bibfnamefont
  {J.}~\bibnamefont {Ye}},\ }\href {\doibase 10.1038/nature12483} {\bibfield
  {journal} {\bibinfo  {journal} {Nature}\ }\textbf {\bibinfo {volume} {501}},\
  \bibinfo {pages} {521} (\bibinfo {year} {2013})}\BibitemShut {NoStop}%
\bibitem [{\citenamefont {Barry}\ \emph {et~al.}(2014)\citenamefont {Barry},
  \citenamefont {McCarron}, \citenamefont {Norrgard}, \citenamefont
  {Steinecker},\ and\ \citenamefont {DeMille}}]{Barry2014}%
  \BibitemOpen
  \bibfield  {author} {\bibinfo {author} {\bibfnamefont {J.~F.}\ \bibnamefont
  {Barry}}, \bibinfo {author} {\bibfnamefont {D.~J.}\ \bibnamefont {McCarron}},
  \bibinfo {author} {\bibfnamefont {E.~B.}\ \bibnamefont {Norrgard}}, \bibinfo
  {author} {\bibfnamefont {M.~H.}\ \bibnamefont {Steinecker}}, \ and\ \bibinfo
  {author} {\bibfnamefont {D.}~\bibnamefont {DeMille}},\ }\href {\doibase
  10.1038/nature13634} {\bibfield  {journal} {\bibinfo  {journal} {Nature}\
  }\textbf {\bibinfo {volume} {512}},\ \bibinfo {pages} {286} (\bibinfo {year}
  {2014})}\BibitemShut {NoStop}%
\bibitem [{\citenamefont {Truppe}\ \emph {et~al.}(2017)\citenamefont {Truppe},
  \citenamefont {Williams}, \citenamefont {Hambach}, \citenamefont {Caldwell},
  \citenamefont {Fitch}, \citenamefont {Hinds}, \citenamefont {Sauer},\ and\
  \citenamefont {Tarbutt}}]{Truppe2017SubDoppler}%
  \BibitemOpen
  \bibfield  {author} {\bibinfo {author} {\bibfnamefont {S.}~\bibnamefont
  {Truppe}}, \bibinfo {author} {\bibfnamefont {H.~J.}\ \bibnamefont
  {Williams}}, \bibinfo {author} {\bibfnamefont {M.}~\bibnamefont {Hambach}},
  \bibinfo {author} {\bibfnamefont {L.}~\bibnamefont {Caldwell}}, \bibinfo
  {author} {\bibfnamefont {N.~J.}\ \bibnamefont {Fitch}}, \bibinfo {author}
  {\bibfnamefont {E.~A.}\ \bibnamefont {Hinds}}, \bibinfo {author}
  {\bibfnamefont {B.~E.}\ \bibnamefont {Sauer}}, \ and\ \bibinfo {author}
  {\bibfnamefont {M.~R.}\ \bibnamefont {Tarbutt}},\ }\href
  {http://arxiv.org/abs/1703.00580} {\  (\bibinfo {year} {2017})},\ \Eprint
  {http://arxiv.org/abs/1703.00580} {arXiv:1703.00580} \BibitemShut {NoStop}%
\bibitem [{\citenamefont {Anderegg}\ \emph {et~al.}(2017)\citenamefont
  {Anderegg}, \citenamefont {Augenbraun}, \citenamefont {Chae}, \citenamefont
  {Hemmerling}, \citenamefont {Hutzler}, \citenamefont {Ravi}, \citenamefont
  {Collopy}, \citenamefont {Ye}, \citenamefont {Ketterle},\ and\ \citenamefont
  {Doyle}}]{Anderegg2017}%
  \BibitemOpen
  \bibfield  {author} {\bibinfo {author} {\bibfnamefont {L.}~\bibnamefont
  {Anderegg}}, \bibinfo {author} {\bibfnamefont {B.}~\bibnamefont
  {Augenbraun}}, \bibinfo {author} {\bibfnamefont {E.}~\bibnamefont {Chae}},
  \bibinfo {author} {\bibfnamefont {B.}~\bibnamefont {Hemmerling}}, \bibinfo
  {author} {\bibfnamefont {N.~R.}\ \bibnamefont {Hutzler}}, \bibinfo {author}
  {\bibfnamefont {A.}~\bibnamefont {Ravi}}, \bibinfo {author} {\bibfnamefont
  {A.}~\bibnamefont {Collopy}}, \bibinfo {author} {\bibfnamefont
  {J.}~\bibnamefont {Ye}}, \bibinfo {author} {\bibfnamefont {W.}~\bibnamefont
  {Ketterle}}, \ and\ \bibinfo {author} {\bibfnamefont {J.}~\bibnamefont
  {Doyle}},\ }\href {http://arxiv.org/abs/1705.10288} {\  (\bibinfo {year}
  {2017})},\ \Eprint {http://arxiv.org/abs/1705.10288} {arXiv:1705.10288}
  \BibitemShut {NoStop}%
\bibitem [{\citenamefont {Monroe}\ \emph {et~al.}(1995)\citenamefont {Monroe},
  \citenamefont {Meekhof}, \citenamefont {King}, \citenamefont {Jefferts},
  \citenamefont {Itano}, \citenamefont {Wineland},\ and\ \citenamefont
  {Gould}}]{Monroe1995}%
  \BibitemOpen
  \bibfield  {author} {\bibinfo {author} {\bibfnamefont {C.}~\bibnamefont
  {Monroe}}, \bibinfo {author} {\bibfnamefont {D.~M.}\ \bibnamefont {Meekhof}},
  \bibinfo {author} {\bibfnamefont {B.~E.}\ \bibnamefont {King}}, \bibinfo
  {author} {\bibfnamefont {S.~R.}\ \bibnamefont {Jefferts}}, \bibinfo {author}
  {\bibfnamefont {W.~M.}\ \bibnamefont {Itano}}, \bibinfo {author}
  {\bibfnamefont {D.~J.}\ \bibnamefont {Wineland}}, \ and\ \bibinfo {author}
  {\bibfnamefont {P.}~\bibnamefont {Gould}},\ }\href {\doibase
  10.1103/PhysRevLett.75.4011} {\bibfield  {journal} {\bibinfo  {journal}
  {Phys. Rev. Lett.}\ }\textbf {\bibinfo {volume} {75}},\ \bibinfo {pages}
  {4011} (\bibinfo {year} {1995})}\BibitemShut {NoStop}%
\bibitem [{\citenamefont {Kerman}\ \emph {et~al.}(2000)\citenamefont {Kerman},
  \citenamefont {Vuleti\ifmmode~\acute{c}\else \'{c}\fi{}}, \citenamefont
  {Chin},\ and\ \citenamefont {Chu}}]{Kerman2000}%
  \BibitemOpen
  \bibfield  {author} {\bibinfo {author} {\bibfnamefont {A.~J.}\ \bibnamefont
  {Kerman}}, \bibinfo {author} {\bibfnamefont {V.}~\bibnamefont
  {Vuleti\ifmmode~\acute{c}\else \'{c}\fi{}}}, \bibinfo {author} {\bibfnamefont
  {C.}~\bibnamefont {Chin}}, \ and\ \bibinfo {author} {\bibfnamefont
  {S.}~\bibnamefont {Chu}},\ }\href {\doibase 10.1103/PhysRevLett.84.439}
  {\bibfield  {journal} {\bibinfo  {journal} {Phys. Rev. Lett.}\ }\textbf
  {\bibinfo {volume} {84}},\ \bibinfo {pages} {439} (\bibinfo {year}
  {2000})}\BibitemShut {NoStop}%
\bibitem [{\citenamefont {Han}\ \emph {et~al.}(2000)\citenamefont {Han},
  \citenamefont {Wolf}, \citenamefont {Oliver}, \citenamefont {McCormick},
  \citenamefont {DePue},\ and\ \citenamefont {Weiss}}]{Han2000}%
  \BibitemOpen
  \bibfield  {author} {\bibinfo {author} {\bibfnamefont {D.-J.}\ \bibnamefont
  {Han}}, \bibinfo {author} {\bibfnamefont {S.}~\bibnamefont {Wolf}}, \bibinfo
  {author} {\bibfnamefont {S.}~\bibnamefont {Oliver}}, \bibinfo {author}
  {\bibfnamefont {C.}~\bibnamefont {McCormick}}, \bibinfo {author}
  {\bibfnamefont {M.~T.}\ \bibnamefont {DePue}}, \ and\ \bibinfo {author}
  {\bibfnamefont {D.~S.}\ \bibnamefont {Weiss}},\ }\href {\doibase
  10.1103/PhysRevLett.85.724} {\bibfield  {journal} {\bibinfo  {journal} {Phys.
  Rev. Lett.}\ }\textbf {\bibinfo {volume} {85}},\ \bibinfo {pages} {724}
  (\bibinfo {year} {2000})}\BibitemShut {NoStop}%
\bibitem [{\citenamefont {Hutzler}\ \emph {et~al.}(2017)\citenamefont
  {Hutzler}, \citenamefont {Liu}, \citenamefont {Yu},\ and\ \citenamefont
  {Ni}}]{Hutzler2017-LightShifts}%
  \BibitemOpen
  \bibfield  {author} {\bibinfo {author} {\bibfnamefont {N.~R.}\ \bibnamefont
  {Hutzler}}, \bibinfo {author} {\bibfnamefont {L.~R.}\ \bibnamefont {Liu}},
  \bibinfo {author} {\bibfnamefont {Y.}~\bibnamefont {Yu}}, \ and\ \bibinfo
  {author} {\bibfnamefont {K.-K.}\ \bibnamefont {Ni}},\ }\href {\doibase
  10.1088/1367-2630/aa5a3b} {\bibfield  {journal} {\bibinfo  {journal} {New J.
  Phys.}\ }\textbf {\bibinfo {volume} {19}},\ \bibinfo {pages} {023007}
  (\bibinfo {year} {2017})}\BibitemShut {NoStop}%
\bibitem [{Note1()}]{Note1}%
  \BibitemOpen
  \bibinfo {note} {The final hyperfine state sensitive detection employs a
  strong beam that is resonant with the $F=2$ to $F'=3$ cycling transition, to
  heat the $F=2$ population out of the trap. Therefore, any atoms which survive
  this heat out procedure are interpreted as having been in the $F=1$
  state.}\BibitemShut {Stop}%
\bibitem [{\citenamefont {Kasevich}\ and\ \citenamefont
  {Chu}(1992)}]{Kasevich1992}%
  \BibitemOpen
  \bibfield  {author} {\bibinfo {author} {\bibfnamefont {M.}~\bibnamefont
  {Kasevich}}\ and\ \bibinfo {author} {\bibfnamefont {S.}~\bibnamefont {Chu}},\
  }\href {\doibase 10.1103/PhysRevLett.69.1741} {\bibfield  {journal} {\bibinfo
   {journal} {Phys. Rev. Lett.}\ }\textbf {\bibinfo {volume} {69}},\ \bibinfo
  {pages} {1741} (\bibinfo {year} {1992})}\BibitemShut {NoStop}%
\bibitem [{\citenamefont {Gr\"obner}\ \emph {et~al.}(2017)\citenamefont
  {Gr\"obner}, \citenamefont {Weinmann}, \citenamefont {Kirilov},\ and\
  \citenamefont {N\"agerl}}]{Grobner2017}%
  \BibitemOpen
  \bibfield  {author} {\bibinfo {author} {\bibfnamefont {M.}~\bibnamefont
  {Gr\"obner}}, \bibinfo {author} {\bibfnamefont {P.}~\bibnamefont {Weinmann}},
  \bibinfo {author} {\bibfnamefont {E.}~\bibnamefont {Kirilov}}, \ and\
  \bibinfo {author} {\bibfnamefont {H.-C.}\ \bibnamefont {N\"agerl}},\ }\href
  {\doibase 10.1103/PhysRevA.95.033412} {\bibfield  {journal} {\bibinfo
  {journal} {Phys. Rev. A}\ }\textbf {\bibinfo {volume} {95}},\ \bibinfo
  {pages} {033412} (\bibinfo {year} {2017})}\BibitemShut {NoStop}%
\bibitem [{Note2()}]{Note2}%
  \BibitemOpen
  \bibinfo {note} {We find a reduction in the scattering rate by a factor of
  $130(20)$, as compared to using an OP resonant with $3^2P_{3/2}$, from which
  the $|2, 2>$ state could always scatter a photon via the excited $|F'=3,
  m_{F'}=3>$ state.}\BibitemShut {Stop}%
\bibitem [{\citenamefont {Wineland}\ and\ \citenamefont
  {Itano}(1979)}]{ItanoWineland1979}%
  \BibitemOpen
  \bibfield  {author} {\bibinfo {author} {\bibfnamefont {D.~J.}\ \bibnamefont
  {Wineland}}\ and\ \bibinfo {author} {\bibfnamefont {W.~M.}\ \bibnamefont
  {Itano}},\ }\href {\doibase 10.1103/PhysRevA.20.1521} {\bibfield  {journal}
  {\bibinfo  {journal} {Phys. Rev. A}\ }\textbf {\bibinfo {volume} {20}},\
  \bibinfo {pages} {1521} (\bibinfo {year} {1979})}\BibitemShut {NoStop}%
\bibitem [{\citenamefont {Dalibard}\ \emph {et~al.}(1992)\citenamefont
  {Dalibard}, \citenamefont {Castin},\ and\ \citenamefont
  {M\o{}lmer}}]{Dalibard1992}%
  \BibitemOpen
  \bibfield  {author} {\bibinfo {author} {\bibfnamefont {J.}~\bibnamefont
  {Dalibard}}, \bibinfo {author} {\bibfnamefont {Y.}~\bibnamefont {Castin}}, \
  and\ \bibinfo {author} {\bibfnamefont {K.}~\bibnamefont {M\o{}lmer}},\ }\href
  {\doibase 10.1103/PhysRevLett.68.580} {\bibfield  {journal} {\bibinfo
  {journal} {Phys. Rev. Lett.}\ }\textbf {\bibinfo {volume} {68}},\ \bibinfo
  {pages} {580} (\bibinfo {year} {1992})}\BibitemShut {NoStop}%
\bibitem [{\citenamefont {Meekhof}\ \emph {et~al.}(1996)\citenamefont
  {Meekhof}, \citenamefont {Monroe}, \citenamefont {King}, \citenamefont
  {Itano},\ and\ \citenamefont {Wineland}}]{Meekhof1996}%
  \BibitemOpen
  \bibfield  {author} {\bibinfo {author} {\bibfnamefont {D.~M.}\ \bibnamefont
  {Meekhof}}, \bibinfo {author} {\bibfnamefont {C.}~\bibnamefont {Monroe}},
  \bibinfo {author} {\bibfnamefont {B.~E.}\ \bibnamefont {King}}, \bibinfo
  {author} {\bibfnamefont {W.~M.}\ \bibnamefont {Itano}}, \ and\ \bibinfo
  {author} {\bibfnamefont {D.~J.}\ \bibnamefont {Wineland}},\ }\href {\doibase
  10.1103/PhysRevLett.76.1796} {\bibfield  {journal} {\bibinfo  {journal}
  {Phys. Rev. Lett.}\ }\textbf {\bibinfo {volume} {76}},\ \bibinfo {pages}
  {1796} (\bibinfo {year} {1996})}\BibitemShut {NoStop}%
\bibitem [{Note3()}]{Note3}%
  \BibitemOpen
  \bibinfo {note} {See supplemental material}\BibitemShut {NoStop}%
\bibitem [{\citenamefont {Grimm}\ \emph {et~al.}(2000)\citenamefont {Grimm},
  \citenamefont {Weidem{\"{u}}ller},\ and\ \citenamefont
  {Ovchinnikov}}]{Grimm2000}%
  \BibitemOpen
  \bibfield  {author} {\bibinfo {author} {\bibfnamefont {R.}~\bibnamefont
  {Grimm}}, \bibinfo {author} {\bibfnamefont {M.}~\bibnamefont
  {Weidem{\"{u}}ller}}, \ and\ \bibinfo {author} {\bibfnamefont {Y.~B.}\
  \bibnamefont {Ovchinnikov}},\ }\href {\doibase 10.1016/S1049-250X(08)60186-X}
  {\bibfield  {journal} {\bibinfo  {journal} {Adv. At. Mol. Opt. Phys.}\
  }\textbf {\bibinfo {volume} {42}},\ \bibinfo {pages} {95} (\bibinfo {year}
  {2000})}\BibitemShut {NoStop}%
\bibitem [{\citenamefont {Colzi}\ \emph {et~al.}(2016)\citenamefont {Colzi},
  \citenamefont {Durastante}, \citenamefont {Fava}, \citenamefont {Serafini},
  \citenamefont {Lamporesi},\ and\ \citenamefont {Ferrari}}]{Colzi2016}%
  \BibitemOpen
  \bibfield  {author} {\bibinfo {author} {\bibfnamefont {G.}~\bibnamefont
  {Colzi}}, \bibinfo {author} {\bibfnamefont {G.}~\bibnamefont {Durastante}},
  \bibinfo {author} {\bibfnamefont {E.}~\bibnamefont {Fava}}, \bibinfo {author}
  {\bibfnamefont {S.}~\bibnamefont {Serafini}}, \bibinfo {author}
  {\bibfnamefont {G.}~\bibnamefont {Lamporesi}}, \ and\ \bibinfo {author}
  {\bibfnamefont {G.}~\bibnamefont {Ferrari}},\ }\href {\doibase
  10.1103/PhysRevA.93.023421} {\bibfield  {journal} {\bibinfo  {journal} {Phys.
  Rev. A}\ }\textbf {\bibinfo {volume} {93}},\ \bibinfo {pages} {023421}
  (\bibinfo {year} {2016})}\BibitemShut {NoStop}%
\end{thebibliography}%
\end{document}